\documentclass[9pt]{article}
\usepackage{amsfonts}
\usepackage{pstricks}
\usepackage{pst-node}
\usepackage{epsfig}
\usepackage{epsfig}
\usepackage{graphicx}
\usepackage[font={footnotesize,it}]{caption}
\usepackage[latin5]{inputenc}
\usepackage[T1]{fontenc}
\usepackage{lipsum}
\usepackage{amsmath}

\begin{document}
                \def\ba{\begin{eqnarray}}
                \def\ea{\end{eqnarray}}
                \def\w{\wedge}
                \def\d{\mbox{d}}
                \def\D{\mbox{D}}

%%%%%%%%%%%%%%%%%%%%%%%%%%%%%%%%%%%% TITLEPAGE %%%%%%%%%%%%%%%%%%%%%%%%%%%%%%
\begin{titlepage}
\title{Neutrino Fields in a Sandwich Gravitational Wave Background}
\author{Tekin Dereli${}^{1,2}$\footnote{tekindereli@maltepe.edu.tr}, Ozay Gurtug${}^{1}$\footnote{ozaygurtug@maltepe.edu.tr}, Mustafa Halilsoy${}^{3}$\footnote{mustafa.halilsoy@emu.edu.tr}, Yorgo Senikoglu${}^{1}$\footnote{yorgosenikoglu@maltepe.edu.tr}, }
\date{%
    ${}^{1}$ \small Department of Basic Sciences, Faculty of Engineering and Natural Sciences, \\Maltepe University, 34857 Maltepe,\.{I}stanbul, Turkey\\%
    ${}^{2}$ \small Emeritus Professor of Physics, Ko\c{c} University, 34450 Sar{\i}yer,\.{I}stanbul, Turkey\\
    ${}^{3}$ \small Department of Physics, Faculty of Arts and Sciences, Eastern Mediterranean University, Famagusta, North Cyprus via Mersin 10, Turkey\\[2ex]%
    \today
}

%\affiliation{Department of Physics, Ko\c{c} University, 34450 Sar{\i}yer-\.{I}stanbul, Turkey }
\maketitle
%\vskip 1cm

%\date{4 September 2019}

%vskip 2cm

\begin{abstract}
\noindent Sandwich gravitational waves are given globally in terms of step functions at the boundaries. Linearized Einstein-Weyl equations are solved exactly in this background in Rosen coordinates. Depending on the geometry and composition of the sandwich wave, the neutrino's energy-momentum redistributes itself. At the test field level, since the background will not change, the neutrino's energy density in particular will show variations between positive and negative extrema when crossing the sandwich wave. This may reveal facts about the weakly interacting neutrinos in cosmology.
\end{abstract}

\vskip 1cm


\end{titlepage}

\newpage
%%%%%%%%%%%%%%%%%%%%%%%%%%%%%%%%%%%% INTRODUCTION %%%%%%%%%%%%%%%%%%%%%%%%%%%%%%

\section{Introduction}
One of the important predictions of the Einstein's theory of relativity is the existence of gravitational waves. In the early stages, there was a belief among the physicists, including Einstein himself, that its experimental evidence is almost hopeless in the foreseeable future \cite{rosen},\cite{einstein-rosen}. In spite of this fact, understanding gravitational waves has kept physicists busy and research along this direction became one of the important topics in theoretical physics \cite{brinkmann},\cite{peres}. Investigation of plane gravitational waves, which are the exact solutions of Einstein and Einstein - Maxwell field equations revealed number of interesting properties \cite{ehlers-kundt}. One of them is the caustic property of plane waves pointed out  by Bondi et al \cite{pirani}. Focusing property of plane waves is studied in the seminal paper of Penrose \cite{penrose}. Later on, it has been understood that the mutual focussing of plane waves are the characteristic property of nonlinear interaction of gravitational waves, i.e.collision of gravitational waves, which leads to the creation of curvature singularities \cite{szekeres},\cite{griffiths0}. All these are the results of the nonlinear essence of the gravitational waves.

We are now confident that there exists gravitational waves and it is no more a subject of solely the theoretical physicists.
Since the discovery of the gravitational waves by the ground based observatories Advanced LIGO and Advanced VIRGO, a new window has opened for observing the universe through the analysis of gravitational waves \cite{abbott et al}. The actual observation of gravitational waves proved that our universe contains propagating gravitational waves that are generated by the merger of binary systems consisting of a pair of black holes or of neutron stars.

Gravitational wave physics is a nonlinear phenomena and thus any information carried by these waves are expected to be distinct when compared to the information obtained from electromagnetic waves. In view of this reality, it is important to investigate, how the fundamental fields when propagating in the background of gravitational waves be affected \cite{gibbons}.
%%%%%%%%%%%%%%%%%%%%%%%%%%
The cases of real scalar fields obeying the Einstein-Klein-Gordon equations or the electromagnetic fields obeying the Einstein-Maxwell equations had been studied a lot.
The case of left-handed neutrino fields that satisfy the coupled Einstein-Weyl equations is less familiar \cite{brill-wheeler},\cite{wainwright},\cite{trim-wainwright},\cite{griffiths1}. In particular, neutrino progressive waves and their collision in gravitational wave spacetimes were studied by Griffiths \cite{griffiths2},\cite{griffiths-podolsky},\cite{dereli-tucker}.
On the other hand,
the propagation of massive Dirac fields through gravitational plane wave were also considered in \cite{bini-ferrari2}. It was found after the scattering of Dirac fields the wave function
gets modified but the spin state remains unchanged.
More recently, Dirac particles are also considered in a pure gravitational sandwich wave \cite{collas}. It was demonstrated that the background pure gravitational sandwich waves may alter the initial spin polarization of the Dirac particles.
These are in line with the results of a previous discussion (Refs.\cite{halilsoy},\cite{albadawi-halilsoy}) where the global form of the gravitational sandwich wave was presented in pure Einstein and Einstein-Maxwell theories. It was proven that a test scalar particle extracts energy while crossing across a  gravitational sandwich wave.
It should be emphasized that in the cases above, both the scalar particles and the Dirac particles are considered as test particles. In other words, the change in the background gravitational wave geometry has not been taken into account; only how the particles/fields are affected by the curved background geometry as they pass through is observed.

As a finite curvature zone, a gravitational sandwich wave may arise due to a very short durational explosion of a gravitational source such as a mininova and since it moves at the speed of light, it traverses a portion of the cosmos. Besides cosmic explosions, our Earth may also pulsate such waves following a strong tremor of an earthquake. Since gravitational interaction is universal, everything including the weakly interacting neutrino becomes sensitive to such gravitational waves. As an electromagnetic wave undergoes a Faraday rotation upon encountering a gravitational wave \cite{halilsoy-gurtug}, a neutrino is also expected to polarize itself accordingly. This however, is a prominent process in case of a full solution of the Einstein-Maxwell system including the background effect. At the test field level, however, the expectations become limited. Even at this level, it is found that the energy distribution shows great variations while crossing the sandwich wave. It remains to be seen whether such variations may trigger the process of neutrino oscillations between different types. Although a neutrino fails to interact with a pure electromagnetic spacetime, it does interact in the presence of a non-zero Weyl curvature.
As shown in Figure 1, a neutrino beam is focused at a distance, upon crossing the sandwich wave. We have a flat-curved-flat sequence of regions in such a picture. The process can be further enhanced by a succession of sandwich waves as long as each wave is prior to the focusing hypersurface. That is, the focusing must be delayed after the passage of the wave. In all theses processes, it should be supplemented that at each boundary the second fundamental form must make a continuous transition in order to avoid the creation of extra surface sources at the boundaries. At the level of Einstein-Maxwell theory, which is our aim in this study, the boundary conditions of O-Brien and Synge are sufficient.

In the present  paper, we consider massless test neutrino fields in the background of sandwich plane waves that may comprise of i) pure gravitational, ii) pure electromagnetic or iii) a mixture of gravitational and electromagnetic sandwich waves.

The plan of the paper is as follows. In section II, the general overview of sandwich gravitational waves in Einstein - Maxwell theory is reviewed. Neutrino field equations are solved in the sandwich gravitational wave background in section III. In section IV, explicit calculations for the energy variations in crossing the sandwich gravitational waves are given. The paper is concluded with a conclusion in section VI.

\bigskip
%%%%%%%%%%%%%%%%%%%%%%%%%%%%%%%  SECTION 2 %%%%%%%%%%%%%%%%%%%%%%%%%%%%%%%%%%%%%%
%\newpage

\section{The Sandwich Gravitational Plane Waves}
Brinkmann's inaugural study \cite{brinkmann} of $pp$-wave spacetimes was first understood in terms of gravitational waves by Peres \cite{peres}. The gravitational or electromagnetic $pp$-wave metric can be given in the Brinkmann form as:
\begin{equation}\label{Brinkmann}
ds^2=2dUdV - 2 H(U,X,Y)dU^2 - dX^2 - dY^2.
\end{equation}
It turns out that the only non-vanishing components of Weyl and Ricci scalars are, respectively,
\begin{equation}
 \psi_4=-\frac{1}{2}(H_{XX}-H_{YY}-2iH_{XY}), \quad
 \phi_{22}=-\frac{1}{2}(H_{XX}+H_{YY}).
\end{equation}
A gravitational wave in $pp$-wave spacetimes \cite{ehlers-kundt,bondi-pirani-robinson} is determined by the single curvature component $\psi_4=|{\psi_4}| e^{i\theta}$, which depends exclusively on the null coordinate $U$ and where $|\psi_4|$ and $\theta$ denote respectively the amplitude and its polarization. In particular, the linearly polarized waves are characterized by constant
$\theta$. On the other hand, the non-zero Ricci curvature component $\phi_{22}$ represents the matter field content of the $pp$-wave.

Sandwich waves constitute a simple extension of shock or step waves that may be expressed in terms of Heaviside step functions that take on a non-zero value over a finite
interval $0\leq U \leq U_0$. The flat Minkowski metric describes the geometry both in the region ahead of $U=0$ and behind  of the wave-front $U=U_0$ in this case.
A global form of a sandwich plane wave metric in Einstein-Maxwell theory can be found in Ref.\cite{halilsoy},  the impulsive gravitational wave restrictions of these were taken into account in \cite{podolsky-vesely} and a generic technique for calculating the dispersion relations of any wave that is propagating in a structure of sandwich wave was provided in \cite{liu-bhatta}. By taking appropriate limits, one can easily construct pure gravitational, or pure electromagnetic or a mixtures of both valid for the finite duration curvature zone. In doing so, a linearly polarized plane sandwich wave is considered to comprise the gravitational
$\psi_4$ and electromagnetic $\phi_{22}$ parts independent of the coordinates $X,Y$ that span the transverse planar wave front. Then the metric function reads
\begin{equation}
  H(U,X,Y)=\frac{1}{2}[(\Theta(U)-\Theta(U-U_0)][a^2(X^2+Y^2)-b^2(X^2-Y^2)],
\end{equation}
where $a$ and $b$ are the electromagnetic and the gravitational parameters, respectively.
It is usually found more convenient to utilize Rosen's metric form \cite{rosen} to demonstrate the transverse feature of such spacetimes, via the coordinate transformation
\ba
U=u, \quad U_0=u_0, \\
V=v + \frac{1}{2}(x^2FF_u+y^2GG_u), \\
X=xF, \quad Y=yG
\ea
so that the metric  takes the form
\begin{equation}\label{Rosen}
  ds^2=2dudv - F(u)^2 dx^2 - G(u)^2 dy^2.
\end{equation}
The functions $F(u)$ and $G(u)$ satisfy the  set of following differential equations
\ba
F_{uu}+A^2(\theta(u)-\theta(u-u_0))F=0, \nonumber \\
G_{uu}+B^2(\theta(u)-\theta(u-u_0))G=0,
\ea
whose generic solution is given by
\ba
F(u)=cos[A(u\theta(u)-(u-u_0)\theta(u-u_0))]-Asin(Au_0)(u-u_0)\theta(u-u_0), \nonumber \\
G(u)=cos[B(u\theta(u)-(u-u_0)\theta(u-u_0))]-Bsin(Bu_0)(u-u_0)\theta(u-u_0).
\ea
We set $A^2=(a^2-b^2)$ and $B^2=(a^2+b^2)$.

\vspace{1mm}
In order to determine the wave profiles, we introduce a null tetrad basis
\begin{equation}
  l=du, \quad n=dv, \quad m=\frac{1}{\sqrt{2}}(Fdx+iGdy) \quad \bar{m}=\frac{1}{\sqrt{2}}(Fdx-iGdy),
\end{equation}
in terms of which the non-vanishing curvature components become
\ba
\psi_4&=&-\frac{F_{uu}}{2F}+\frac{G_{uu}}{2G} = -b^2(\theta(u)-\theta(u-u_0)), \nonumber \\
\phi_{22}&=&-\frac{F_{uu}}{2F}-\frac{G_{uu}}{2G}=a^2(\theta(u)-\theta(u-u_0)).
\ea

\vspace{2mm}
In this paper, the gravitational wave background will be studied for three cases corresponding,  firstly, to a pure gravitational, or secondly to a pure electromagnetic or finally to a mixture of  gravitational and electromagnetic sandwich waves. In all these cases the relevant quantities  can be obtained from the obtained generic solution by going to the appropriate limits.

The general scheme for considering such a gravitational sandwich wave geometry is illustrated in Figure 1. The region $u<0$ is a flat Minkowski region where the metric functions $F(u)=G(u)=1$ and the corresponding metric is given by
\begin{equation}
  ds^2=2dudv - dx^2 - dy^2.
\end{equation}
Region II, which is characterized when $0\leq u \leq u_0$ is a non-trivial curved region that contains a sandwich wave with an endowed metric given by
\begin{equation}
  ds^2=2dudv - F(u)^2dx^2 - G(u)^2dy^2.
\end{equation}
We remind here that the metric functions $F(u)$ and $G(u)$ correspond to the considered background geometry of the specific sandwich wave.
Finally the
Region III, described by  $u > u_0$ is again a flat region. However, this is not explicit as in the case of Region I. In fact the metric in Region III is given by
\begin{equation}\label{region3}
  ds^2=2dudv - \tilde{F}(u)^2dx^2 - \tilde{G}(u)^2dy^2,
\end{equation}
The flat character of the metric (\ref{region3}), can be shown after  the following transformation is applied
\ba
U&=&u, \nonumber \\
X&=&x\tilde{F}, \quad
Y=y\tilde{G}, \nonumber \\
V&=&v+\frac{1}{2}x^{2}\tilde{F}\tilde{F}_u+\frac{1}{2}\tilde{G}\tilde{G}_u y^{2}
\ea
that yields
\begin{equation}
ds^{2}=2dUdV-dX^{2}-dY^{2}.
\end{equation}

%\begin{center}
%\scalebox{0.45}{\includegraphics{swgeometry1}}
%\end{center}

\begin{figure}
  \centering
  \includegraphics[width=1.00 \textwidth]{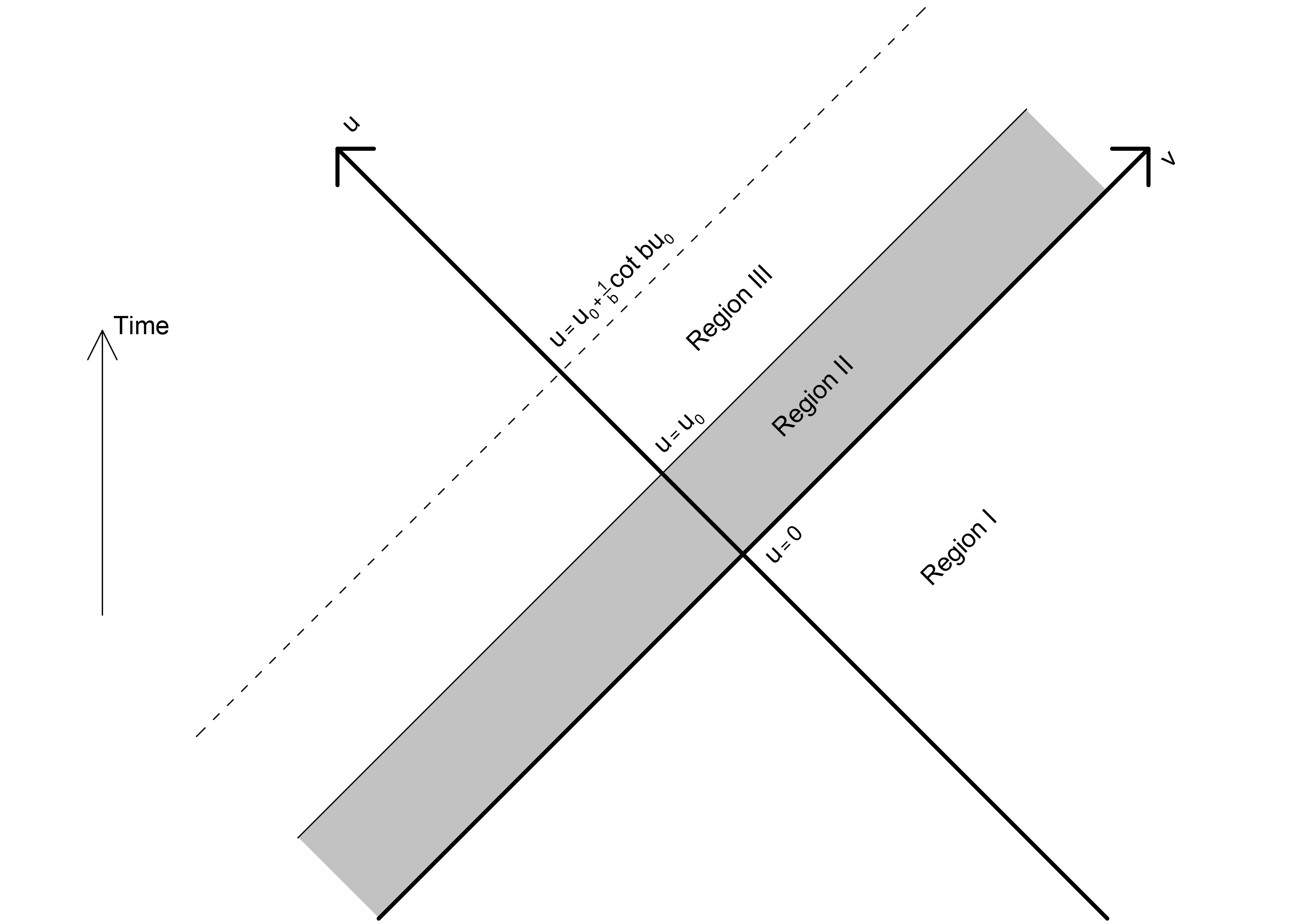}
  \caption{The general scheme of the gravitational sandwich wave geometry. Region I is the flat Minkowski region, which is the front of sandwich wave. Region II is the curved region that contains sandwich wave. Region III is the back of the sandwich wave. The dashed line indicates the singularity where the light is focussed to a line for pure gravitational sandwich wave.}
\end{figure}

These three independent regions, two flat ones outside and one curved inside the finite-duration plane fronted wave region, must be patched together by employing appropriate junction conditions. For colliding plane electromagnetic waves in Einstein-Maxwell theory, Bell and Szekeres \cite{bell-szekeres} have shown that the appropriate junction conditions are those of O'Brien and Synge that require the continuity of the metric $g_{\mu\nu}$ and $g^{ij}g_{ij,u}$, thus allowing us to avoid any source term across the boundaries.
It has been shown before \cite{halilsoy} that the O'Brien and Synge junction conditions are satisfied across the two boundaries $u=0$ and $u=u_{0}$, for the global sandwich wave solution in Einstein-Maxwell theory.

Another important point to be clarified concerns the question of singularities in Regions II and III. In order to avoid a singularity in Region II, it is assumed that $bu<\frac{\pi}{2}$ and $au<\frac{\pi}{2}$, for pure gravitational and electromagnetic sandwich waves, respectively. On the other hand, for a pure gravitational sandwich wave in particular, the apparent singularity in Region III yields
\begin{equation}
u=u_{0}+\frac{1}{b}cot(bu_{0}).
\end{equation}
This is actually a quasi-regular singularity \cite{griffiths1}, which implies the non-existence of complete space-like hypersurfaces in plane wave spacetimes that is adequate for the global specification of Cauchy data. As demonstrated above, by the coordinate transformation (15),  the Region III is flat and hence free of singularity.\\

It is also remarkable to note that from the classical point of view, there seems to be no obstacle against  shrinking the width of the sandwich wave to zero $(u_{0}=0)$. However, from the view point of quantum mechanics, there is a lower bound on $u_{0}$. In order to obtain this lower limit, we appeal to the uncertainty principle:
\begin{equation}
\Delta U \Delta p_{u}\geq\hbar,
\end{equation}
where $\Delta U=u_{0}$ and $\Delta p_{u}$ is the shift in the conjugate momentum. By using the geodesic Lagrangian $L$ in which the canonical momenta are denoted by $p_{u}=\frac{\partial L}{\partial \dot{u}}$ etc., we obtain a lower bound
\begin{equation}
u_{0}\geq \frac{2\alpha \hbar}{p_{2}^{2}tan^{2}(bu_{0})-p_{1}^{2}tanh^{2}(bu_{0})},
\end{equation}
Thus, quantum theory imposes a lower limit to the width of a sandwich wave, provided $0 \leq bu_0 \leq \frac{\pi}{2}$.
\bigskip

\section{Neutrino Equation in a Sandwich Wave Background}

\noindent We describe a left-handed massless test neutrino field  by
 the 2-component spinor
\ba
\Phi = \left ( \begin{array}{c} \varphi_{1} \\ \varphi_{2} \end{array}  \right )
\ea
that satisfies the Weyl equation\footnote{Adding a Majorana mass to the Weyl neutrino was considered nevertheless the equations are not integrable.}
\ba
\sigma^{a} \nabla_{X_{a}} \Phi =0.
\ea
Here $\sigma^a :\{I ,\sigma^1, \sigma^2, \sigma^3 \}$ are the Pauli matrices, $\nabla_{X_{a}}$'s denote covariant derivatives relative to an orthonormal
co-frame $\{e^a\}$, so that the covariant exterior derivative operator $\nabla = e^a \nabla_{X_{a}}$. The (odd-Grassmann) components $\varphi_{1}$ and $\varphi_{2}$
are  taken as complex valued functions of all coordinates $\{u,v,x,y\}$.
The corresponding orthonormal components of the symmetrized energy-momentum tensor are given by
\ba \label{stress-energy}
T_{ab}[\Phi]  = \frac{i}{4} \left ( \Phi^{\dagger} \sigma_a \nabla_{X_{b}} \Phi + \Phi^{\dagger} \sigma_b \nabla_{X_{a}} \Phi
-  \nabla_{X_{a}}\Phi^{\dagger} \sigma_b \Phi - \nabla_{X_{b}}\Phi^{\dagger}  \sigma_a \Phi  \right ).
\ea

\noindent Now we consider the gravitational plane wave metric
\ba
ds^2 = 2du dv - F(u)^{2} dx^2 - G(u)^{2} dy^2
\ea
and work out the Weyl equation in this background spacetime. It reduces to the following system of first order differential equations
\begin{eqnarray}
\left ( \frac{\partial}{\partial v} \right ) \varphi_2 +  \frac{1}{\sqrt{2}} \left ( \frac{1}{F} \frac{\partial}{\partial x}  - \frac{i}{G} \frac{\partial}{\partial y}  \right ) \varphi_1 &=& 0, \\
\left (\frac{\partial}{\partial u}  +\frac{1}{2} ( \frac{F_u}{F} +  \frac{G_u}{G}  ) \right ) \varphi_1+  \frac{1}{\sqrt{2}} \left ( \frac{1}{F} \frac{\partial}{\partial x} + \frac{i}{G} \frac{\partial}{\partial y}  \right ) \varphi_2 &=& 0 .
\end{eqnarray}
A family of  exact solutions is given by
\begin{eqnarray}
\varphi_1 &=& \frac{\xi}{\sqrt{FG}} e^{iK(u)} e^{i(\alpha v - p_1 x - p_2 y)} ,      \\
\varphi_2 &=& \frac{\xi}{\sqrt{FG}} \frac{1}{\alpha \sqrt{2}} \left ( \frac{p_1}{F} -i \frac{p_2}{G} \right )e^{iK(u)} e^{i(\alpha v - p_1 x - p_2 y)} ,
\end{eqnarray}
where $\alpha,p_1, p_2$ are constants of momenta and the common factor $\xi$ is a complex odd-Grassmann constant. The phase function $K(u)$ will be determined in each region separately by evaluating the integral
\ba
K(u) = \frac{1}{2 \alpha} \int_{0}^{u} \left ( \frac{p_1^2}{F(u^{\prime})^{2}} + \frac{p_2^2}{G(u^{\prime})^{2}}  \right ) du^{\prime}.
\ea
\subsection{Neutrino in a pure Gravitational Sandwich Wave}
The metric functions describing a pure gravitational sandwich is obtained by setting $a=0$ in Eq(9), which yields
\ba
F(u)&=&\left\{
    \begin{array}{ll}
    1 , \quad u<0, & Region \hspace{2mm}I \\
    cosh(bu), \quad 0<u<u_0, & Region \hspace{2mm}I \\
    \alpha_0+\beta_0u, \quad u_0<u, & Region \hspace{2mm}III \\
\end{array}\right. \\
G(u)&=&\left\{
    \begin{array}{ll}
    1 , \quad u<0, & Region \hspace{2mm}I \\
    cos(bu), \quad  0<u<u_0, & Region \hspace{2mm}II \\
    \gamma_0-\tau_0u, \quad u_0<u, & Region \hspace{2mm}III \\
    \end{array}\right.
\ea
where
\ba
\alpha_{0}&=&cosh(bu_{0})-bu_{0}sinh(bu_{0}), \nonumber \\
\beta_{0}&=&bsinh(bu_{0}), \nonumber \\
\gamma_{0}&=&cos(bu_{0})+bu_{0}sin(bu_{0}), \nonumber \\
\tau_{0}&=&bsin(bu_{0}). \nonumber
\ea

We find an exact solution of the neutrino field equations in the background of a pure gravitational sandwich wave as
\ba
\tilde{\varphi_1} = \frac{\varphi_1}{\xi} e^{-i(\alpha v - p_1 x - p_2 y)}= \left\{
    \begin{array}{ll}
        e^{iK(u)} & Region \hspace{2mm}I \\
        \frac{e^{iK(u)}}{\sqrt{cosh(bu)cos(bu)}} & Region \hspace{2mm}II \\
        \frac{e^{iK(u)}}{\sqrt{(\alpha_0+\beta_0u)(\gamma_0-\tau_0u)}} & Region \hspace{2mm}III \\
    \end{array}\right.
\ea

\ba
\tilde{\varphi_2} &=& \sqrt{2}\alpha\frac{\varphi_2}{\xi} e^{-i(\alpha v - p_1 x - p_2 y)}\\ \nonumber
&=& \left\{
    \begin{array}{ll}
        (p_1-ip_2)e^{iK(u)} & Region \hspace{2mm}I \\
        (\frac{p_1}{cosh(bu)}-i\frac{p_2}{cos(bu)})\frac{e^{iK(u)}}{\sqrt{cosh(bu)cos(bu)}} & Region \hspace{2mm}II \\
        (\frac{p_1}{\alpha_0+\beta_0u}-i\frac{p_2}{\gamma_0-\tau_0u)})\frac{e^{iK(u)}}{\sqrt{(\alpha_0+\beta_0u)(\gamma_0-\tau_0u)}} & Region \hspace{2mm}III \\
    \end{array}\right.
\ea
In the solutions above, the phase function for each region is calculated as
\ba
2\alpha K(u)= \left\{
    \begin{array}{ll}
        (p_1^2+p_2^2)u+c_1 & Region \hspace{2mm}I \\
        \frac{p_1^2}{b}tanh(bu)+\frac{p_2^2}{b}tan(bu)+c_2 & Region \hspace{2mm}II \\
        \frac{p_2^2}{\tau_0(\gamma_0-\tau_0u)}-\frac{p_1^2}{\beta_0(\alpha_0+\beta_0u)}+c_3 & Region \hspace{2mm}III, \\
    \end{array}\right.
\ea
where $c_1$, $c_2$ and $c_3$ are the integration constants to be fixed by requiring the continuity of the neutrino fields across each boundary.
The calculation for the constants yields
\ba
c_1=c_2=0 \quad and \quad c_3=\frac{p_1^2}{b}coth(bu_0)-\frac{p_2^2}{b}cot(bu_0).
\ea

\subsection{Neutrino in a pure Electromagnetic Sandwich Wave}
 The metric functions corresponding to a pure electromagnetic sandwich wave is obtained if we set $b=0$ in Eq(9); in this particular limit, the metric functions are
 \ba
 F(u)=G(u)=\left\{
    \begin{array}{ll}
    1 , \quad u<0, & Region \hspace{2mm}I \\
    cos(au) , \quad 0<u<u_0, & Region \hspace{2mm}II \\
    \bar{\alpha}_0-\bar{\beta}_0u , \quad u_0<u, & Region \hspace{2mm}III \\
    \end{array}\right.
 \ea
where
\ba
\bar{\alpha}_{0}&=&cos(au_{0})+au_{0}sin(au_{0}) \nonumber \\
\bar{\beta}_{0}&=&asin(au_{0}).
\ea

\noindent
In the background of a pure electromagnetic sandwich wave, an exact solution of the neutrino field equations is
\ba
\tilde{\varphi_1} = \frac{\varphi_1}{\xi} e^{-i(\alpha v - p_1 x - p_2 y)}= \left\{
    \begin{array}{ll}
        e^{iK(u)} & Region \hspace{2mm}I \\
        \frac{e^{iK(u)}}{cos(au)} & Region \hspace{2mm}II \\
        \frac{e^{iK(u)}}{(\bar{\alpha}_0-\bar{\beta}_0u)} & Region \hspace{2mm}III \\
    \end{array}\right.
\ea

\ba
\tilde{\varphi_2} &=& \sqrt{2}\alpha\frac{\varphi_2}{\xi} e^{-i(\alpha v - p_1 x - p_2 y)}\\ \nonumber
&=& \left\{
    \begin{array}{ll}
        (p_1-ip_2)e^{iK(u)} & Region \hspace{2mm}I \\
        \frac{p_1-ip_2}{cos^2(au)}e^{iK(u)} & Region \hspace{2mm}II \\
        \frac{p_1-ip_2}{(\bar{\alpha}_0-\bar{\beta}_0u)^2}e^{iK(u)} & Region \hspace{2mm}III \\
    \end{array}\right.
\ea
In the solutions above, the phase function for each region is calculated as
\ba
2\alpha K(u)= \left\{
    \begin{array}{ll}
        (p_1^2+p_2^2)u+\bar{c}_1 & Region \hspace{2mm}I \\
        \frac{p_1^2+p_2^2}{a}tan(au)+\bar{c}_2 & Region \hspace{2mm}II \\
        \frac{p_1^2+p_2^2}{\bar{\beta}_0(\bar{\alpha}_0-\bar{\beta}_0u)}+\bar{c}_3 & Region \hspace{2mm}III, \\
    \end{array}\right.
\ea
where $\bar{c}_1$, $\bar{c}_2$ and $\bar{c}_3$ are the integration constants to be fixed by considering the continuity of the neutrino fields across each boundary.
The calculation for the constants yields
\ba
\bar{c}_1=\bar{c}_2=0 \quad and \quad \bar{c}_3=-\frac{p_1^2+p_2^2}{a}cot(au_0).
\ea

\subsection{Neutrino in a mixture of gravitational and electromagnetic sandwich waves}
The specific case of a mixture of gravitational and electromagnetic sandwich wave is obtained when $a=b$ in Eq(9), which yields
\ba
F(u)&=&1, \quad everywhere \nonumber \\
G(u)&=&\left\{
    \begin{array}{ll}
    1 , \quad u<0, & Region \hspace{2mm}I \\
    cos(\sqrt{2}au) , \quad 0<u<u_0, & Region \hspace{2mm}II \\
    \tilde{\alpha}_0-\tilde{\beta}_0u , \quad u_0<u, & Region \hspace{2mm}III \\
    \end{array}\right.
\ea
where
\ba
\tilde{\alpha}_{0}&=&cos(\sqrt{2}au_{0})+\sqrt{2}au_{0}sin(\sqrt{2}au_{0}), \nonumber \\
\tilde{\beta}_{0}&=&\sqrt{2}asin(\sqrt{2}au_{0}).
\ea

\noindent
In the background of a mixture of electromagnetic and gravitational sandwich waves, an exact solution of the neutrino field equations can be given by
\ba
\tilde{\varphi_1} = \frac{\varphi_1}{\xi} e^{-i(\alpha v - p_1 x - p_2 y)}= \left\{
    \begin{array}{ll}
        e^{iK(u)} & Region \hspace{2mm}I \\
        \frac{e^{iK(u)}}{\sqrt{cos(\sqrt{2}au)}} & Region \hspace{2mm}II \\
        \frac{e^{iK(u)}}{\sqrt{(\tilde{\alpha}_0-\tilde{\beta}_0u)}} & Region \hspace{2mm}III \\
    \end{array}\right.
\ea

\ba
\tilde{\varphi_2} &=& \sqrt{2}\alpha\frac{\varphi_2}{\xi} e^{-i(\alpha v - p_1 x - p_2 y)}\\ \nonumber
&=& \left\{
    \begin{array}{ll}
        (p_1-ip_2)e^{iK(u)} & Region \hspace{2mm}I \\
        \frac{e^{iK(u)}}{\sqrt{cos(\sqrt{2}au)}}(p_1-\frac{ip_2}{cos(\sqrt{2}au)}) & Region \hspace{2mm}II \\
        \frac{e^{iK(u)}}{\sqrt{(\tilde{\alpha}_0-\tilde{\beta}_0u)}}(p_1-\frac{ip_2}{\tilde{\alpha}_0-\tilde{\beta}_0u}) & Region \hspace{2mm}III \\
    \end{array}\right.
\ea
In the solutions above, the phase function for each region is calculated as
\ba
2\alpha K(u)= \left\{
    \begin{array}{ll}
        (p_1^2+p_2^2)u+\tilde{c}_1 & Region \hspace{2mm}I \\
        p_1^2u+\frac{p_2^2}{\sqrt{2}a}tan(au)+\tilde{c}_2 & Region \hspace{2mm}II \\
        p_1^2u+\frac{p_2^2}{\tilde{\beta}_0(\tilde{\alpha}_0-\tilde{\beta}_0u)}+\tilde{c}_3 & Region \hspace{2mm}III, \\
    \end{array}\right.
\ea
where $\tilde{c}_1$, $\tilde{c}_2$ and $\tilde{c}_3$ are the integration constants are fixed by considering the continuity of the neutrino fields accross each boundary.
The calculation yields
\ba
\tilde{c}_1=\tilde{c}_2=0 \quad and \quad \tilde{c}_3=-\frac{p_2^2}{\sqrt{2}a}cot(\sqrt{2}au_0).
\ea

\section{Energy Considerations}

\noindent  In order to understand the nature of the impact of a sandwich gravitational wave spacetime on a test neutrino field,
one should  analyze the neutrino plane wave solutions given above. Working  out the components of the neutrino stress-energy-momentum tensor (\ref{stress-energy})  is critical for that purpose. The orthonormal components of the neutrino stress-energy-momentum tensor $T_{ab}$ can be used to define
the neutrino stress-energy 3-forms $\tau_{a} = T_{ab} *e^b$  which in turn may be decomposed for  physical interpretation into a (3+1) form structure relative to time-like inertial observer curves. The explicit form of the inertial decomposition of the neutrino stress-energy-momentum  tensor is worked out in the Appendix.
Here we give the change in the neutrino energy density function as the incident test neutrino in Region I passes through Region II into Region III.

We have in a pure gravitational sandwich wave background,
\ba
\Delta \rho &=& \rho_{out} - \rho_{in} \nonumber \\
&=&\frac{\xi\xi^*}{\sqrt{2}\alpha^2}\Bigg[\frac{1}{4\alpha}\Big(p_2^2tan^2(bu_0)-p_1^2tanh^2(bu_0)\Big) \nonumber \\ & & \times \Big(4\alpha^3 + p_1^2(sech^2(bu_0))+p_2^2(sec^2(bu_0)+1) \Big)\nonumber \\
& &-\frac{b}{2}p_1p_2sech(bu_0)sec(bu_0)\Bigg(tanh(bu_0)+tan(bu_0) \Bigg) \Bigg] ,
\ea
while in a pure electromagnetic sandwich wave background,
\ba
\Delta \rho &=& \rho_{out} - \rho_{in} \nonumber \\
&=&\frac{\xi\xi^*}{\sqrt{2}\alpha^2a^2cos^2(au_0)} \Bigg[\alpha\Big(p_1^2+p_2^2\Big)\Big(1+a^2sin^2(au_0)\Big) \nonumber \\
&& + \frac{(p_1^2+p_2^2)^2}{4\alpha}\Big(sec^2(au_0)-a^2cos^2(au_0) \Big) \Bigg] .
\ea
On the other hand, in a mixed gravitational and electromagnetic sandwich wave background we have
\ba
\Delta \rho &=& \rho_{out} - \rho_{in} \nonumber \\
&=&\frac{\xi\xi^*}{\sqrt{2}\alpha^2} \Bigg[\frac{p_2^2\alpha}{2}\Big(sec^2(\sqrt{2}au_0)(1+\frac{1}{2a^2}-2)\Big) \nonumber \\
&+&\frac{1}{4\alpha}\Big( (p_1^2+\frac{p_2^2sec^2(\sqrt{2}au_0)}{2a^2})(p_1^2+p_2^2sec^2(\sqrt{2}au_0))-(p_1^2+p_2^2)^2\Big) \nonumber \\ &-&\frac{ap_1p_2}{\sqrt{2}cos(\sqrt{2}au_0)tan(\sqrt{2}au_0)}\Bigg].
\ea
%%%%%%%%%%%%%%%%%%%%%%%%%%%%%   CONCLUSION    %%%%%%%%%%%%%%%%%%%%%%%%%%%%%%%%%%

\bigskip

\section{Conclusion}

\noindent Here, we studied a test neutrino field in a gravitational sandwich wave background space-times.
The space-time metric in general is given in Rosen coordinates $u,v,x.y.$
A sandwich wave is a special kind of a gravitational wave
whose curvature is non-zero only over a finite region $0 \leq u \leq u_0$.
We dealt separately with all the cases of a pure gravitational sandwich wave, an electromagnetic sandwich wave or a mixture of gravitational and electromagnetic sandwich waves.
The Minkowski metric prevails both in the front and at the back of such sandwich waves. Exact background solutions of the Weyl equation is found in these three regions.
Then they are patched together across the boundaries by the O'Brien-Synge junction conditions. We explicitly demonstrated that in all these cases variations occur in the neutrino energy density when crossing the sandwich gravitational wave. Depending on parameter values, a test neutrino undergoes oscillations in energy while passing through a sandwich gravitational wave. The phase of the neutrino field would also be shifted. Both of these effects are interesting on their own right and should be further investigated whether they allow present-day detection.

Even though the standard electroweak theory that involves three neutrino species was put forward under the assumption of massless, left-chiral neutrinos, it is well-known that
it is possible for neutrinos of one species to change into another as they propagate in space. Such so-called neutrino oscillations are possible only if they are
massive with non-vanishing mass differences between the species. In order to comprehend the flavor oscillation and mass differences, we need to consider at least two species of neutrinos in our model, which will be inquired in a further study.  The fact that the neutrino species do, at least the atmospheric ones, indeed oscillate was observationally confirmed at the turn of 2000's. See e.g.Ref.\cite{senjanovic}. We have shown here that test neutrinos can exchange energy with sandwich gravitational waves. This could be an indication  that in a wider context with more than a single neutrino species, a closer inspection of the gravitational interactions of various types of neutrinos might contribute to their mass differences, and hence
trigger neutrino oscillations.

\bigskip
%\newpage

\section{Appendix}
It is always possible to choose a local frame $e^0=dt$ that is dual to the unit time-like tangent vector field of an inertial observer.
Let $M_3$ be the space-like submanifold with respect to $g$ that contains any tangent vector of the orthogonal $3$-frame\cite{dereli-tucker}.
Once $\tau_a$ is decomposed into its constituents in this frame as
\ba
\tau^0&=&j \w dt + \rho \nonumber \\
\tau^i&=&\mu^i \w dt +G^i,
\ea
one can delineate the physical meaning of the various components of the stress-energy-momentum tensor. In particular,
$j$ is the local energy current density 2-form and $\rho$ is the associated energy density 3-form. The orthonormal $i$-component {\sl force}  $F^i$ per unit area is defined by $\mu^i$, which are nothing but the (Maxwell) stress 2-forms on $M_3$. For non-equilibrium configurations, $G^i$'s measure the corresponding 3-momentum density 3-forms on $M_3$.
For our generic test neutrino solution above, we worked out
the energy density 3-form
\ba
\rho&=&\frac{\xi\xi^*}{\sqrt{2}\alpha^2}\Big[K_u \alpha^2+\alpha^3 + \frac{(K_u+\alpha)}{2}(\frac{p_1^2}{F^2}+\frac{p_2^2}{G^2}) \nonumber \\
& &-\frac{p_1p_2}{2FG}(\frac{F_u}{F}-\frac{G_u}{G})\Big]dx \w dy \w dz;
\ea
the local energy current density 2-form
\ba
j&=&\frac{\xi\xi^*}{\sqrt{2}\alpha^2}\Big(K_u\alpha^2-\alpha^3+\frac{(K_u-\alpha)}{2}(\frac{p_1^2}{F^2}+\frac{p_2^2}{G^2})-\frac{p_1p_2}{2FG}(\frac{F_u}{F}-\frac{G_u}{G})\Big)dx \w dy \nonumber \\
&+&\frac{\xi\xi^*}{2FG^2\alpha^2}\Big( \alpha G_up_1 +p_2F(\frac{p_1^2}{F^2}+\frac{p_2^2}{G^2}+2\alpha^2) \Big)dx \w dz \nonumber \\
&+&\frac{\xi\xi^*}{2F^2G\alpha^2}\Big( \alpha F_up_2 - p_1G(\frac{p_1^2}{F^2}+\frac{p_2^2}{G^2}+2\alpha^2) \Big)dy \w dz;
\ea
 the components of the (Maxwell) stress density 2-forms
\ba
\mu^1&=&\frac{\xi\xi^*\sqrt{2}}{\alpha}\Big( (K_u-\alpha)\frac{p_1}{F} + \frac{p_2G_u}{2G^2} \Big)dx \w dy
+\frac{\xi\xi^*}{G\alpha}\Big(\frac{G_u}{G}+\frac{2p_1p_2}{FG}\Big) dx \w dz \nonumber \\
&& + \frac{\xi\xi^*}{F\alpha}\Big( -\frac{2p_1^2}{F^2} \Big) dy \w dz,
\ea
\ba
\mu^2&=&\frac{\xi\xi^*\sqrt{2}}{\alpha}\Big( (K_u-\alpha)\frac{p_2}{G} - \frac{p_1F_u}{2F^2} \Big)dx \w dy
+\frac{\xi\xi^*}{G\alpha}\Big( \frac{2p_2^2}{G^2} \Big) dx \w dz  \nonumber \\
&& + \frac{\xi\xi^*}{F\alpha}\Big(-\frac{F_u}{F}+\frac{2p_1p_2}{FG}\Big) dy \w dz, \nonumber
\ea
\ba
\mu^3&=&\frac{\xi\xi^*}{\sqrt{2}\alpha^2}\Big(K_u\alpha^2-\alpha^3-\frac{(K_u-\alpha)}{2}(\frac{p_1^2}{F^2}+\frac{p_2^2}{G^2})+\frac{p_1p_2}{2FG}(\frac{F_u}{F}-\frac{G_u}{G})\Big)dx \w dy \nonumber \\
&+&\frac{\xi\xi^*}{2FG^2\alpha^2}\Big(-\alpha G_up_1 - p_2F(\frac{p_1^2}{F^2}+\frac{p_2^2}{G^2}-2\alpha^2) \Big)dx \w dz \nonumber \\
&+&\frac{\xi\xi^*}{2F^2G\alpha^2}\Big(-\alpha F_up_2 + p_1G(\frac{p_1^2}{F^2}+\frac{p_2^2}{G^2}-2\alpha^2) \Big)dy \w dz; \nonumber
\ea
and finally the components of the  3-momentum density 3-forms
\ba
G^1&=&\frac{\xi\xi^*\sqrt{2}}{\alpha}\Big( (K_u+\alpha)\frac{p_1}{F} + \frac{p_2G_u}{2G^2} \Big)dx \w dy \w dz,
\ea
\ba
G^2&=&\frac{\xi\xi^*\sqrt{2}}{\alpha}\Big( (K_u+\alpha)\frac{p_2}{G} - \frac{p_1F_u}{2F^2} \Big)dx \w dy \w dz, \nonumber
\ea
\ba
G^3&=&\frac{\xi\xi^*}{\sqrt{2}\alpha^2}\Big(K_u \alpha^2+\alpha^3 - \frac{(K_u+\alpha)}{2}(\frac{p_1^2}{F^2}+\frac{p_2^2}{G^2})+\frac{p_1p_2}{2FG}(\frac{F_u}{F}-\frac{G_u}{G})\Big)dx \w dy \w dz. \nonumber
\ea

\bigskip

\section{Acknowledgement}
One of us (T.D.) thanks the Turkish Academy of Sciences (TUBA) for partial support.

\end{document}